\documentclass[twocolumn,secnumarabic,amssymb,groupaddress,nofootinbib,aps,pra,showkeys]{revtex4-1}

\usepackage{amsthm}
\usepackage{amsmath}
\usepackage{amssymb}
\usepackage{color}
\usepackage{adjustbox}
\usepackage{graphicx}
\usepackage{hyperref}
\usepackage{multirow}
\usepackage{pifont}
\usepackage{textcomp}

\begin{document}
\title{High-quality photonic entanglement for wavelength-multiplexed quantum communication based on a silicon chip}

\author{Florent Mazeas$^{1,\dagger}$, Michele Traetta$^{1,2,\dagger}$, Marco Bentivegna$^{1,3,\dagger}$, Florian Kaiser$^1$, Djeylan Aktas$^1$, Weiwei Zhang$^4$, Carlos Alonso Ramos$^4$, Lutfi-Arif Bin-Ngah$^1$, Tommaso Lunghi$^1$, \'Eric Picholle$^{1}$, Nadia Belabas-Plougonven$^{5}$, Xavier Le Roux$^4$, \'Eric Cassan$^4$, Delphine Marris-Morini$^4$, Laurent Vivien$^4$, Gr\'egory Sauder$^1$, Laurent Labont\'e$^1$,}
\email{laurent.labonte@unice.fr}
\author{S\'ebastien Tanzilli$^1$}

\affiliation{$^1$Universit\'{e} C\^ote d'Azur, CNRS, Laboratoire de Physique de la Mati\`ere Condens\'{e}e, France\\
$^2$Dipartimento di Elettronica Informatica e Sistemistica, University of Bologna, Bologna, 40136, Italy\\
$^3$Istituto Nazionale di Fisica della Materia, Dipartimento di Fisica, University "La Sapienza", Roma, 00185, Italy\\
$^4$Centre de Nanosciences et de Nanotechnologies, CNRS, Universit\'{e} Paris-Sud, Universit\'{e} Paris-Saclay, C2N~$-$~Marcoussis, 91460 Marcoussis, France\\
$^5$Centre de Nanosciences et de Nanotechnologies, CNRS, Universit\'{e} Paris-Sud, Universit\'{e} Paris-Saclay, C2N~$-$~Orsay, 91405 Orsay cedex, France\\
$^\dagger$These authors contributed equally to this work
}
\date{\today}

\keywords{quantum communication, silicon photonics, multiplexing, nonlinear optics, four-wave mixing}

\begin{abstract}
We report an efficient energy-time entangled photon-pair source based on four-wave mixing in a CMOS-compatible silicon photonics ring resonator. Thanks to suitable optimization, the source shows a large spectral brightness of 400\,pairs of entangled photons /s/MHz for $\rm 500\,\mu W$ pump power. Additionally, the resonator has been engineered so as to generate a frequency comb structure compatible with standard telecom dense wavelength division multiplexers. We demonstrate high-purity energy-time entanglement, \textit{i.e.}, free of photonic noise, with near perfect raw visibilities ($>$~98\%) between various channel pairs in the telecom C-band. Such a compact source stands as a path towards more complex quantum photonic circuits dedicated to quantum communication systems.
\end{abstract}

\maketitle



\section{Introduction}

Integrated photonics plays a major role in the development of classical information technologies~\cite{IntPhot}. Significant efforts have been devoted to multiplexing operations for addressing the challenge of high-capacity and high-speed telecommunication links. Among the physical observables that can be exploited to multiplex data streams, one of the most mature techniques for long-distance communication lies in exploiting dense wavelength-division multiplexing (DWDM) thanks to its natural immunity to propagation disturbances (polarization mode and chromatic dispersions). On the quantum communication side, current integrated circuits operate almost exclusively in the single mode regime, and typically under-exploit the tremendous abilities offered by standard fiber optical technologies. In the perspective of improving quantum key distribution (QKD) systems, a key step lies in the development of robust and reliable devices, showing straightforward compatibility with telecom standards and, consequently, with multimodal operations in the spectral domain~\cite{Lim10, Ngah_LPR15, Meany_LPR14,aktas:hal-01054242,reimer_integrated_2014}.

Silicon-on-insulator (SOI) represents one of the most promising technological platforms, offering high integration density and CMOS compatibility~\cite{harris_integrated_2014}. Recently, SOI has also proven its suitability for the generation of correlated photon pairs through spontaneous four-wave mixing (SFWM)~\cite{azzini_classical_2012,jiang_silicon-chip_2015,grassani_energy_2016,Wakabayashi:15, PhysRevApplied.4.021001,silverstone_qubit_2015,silverstonej._w._-chip_2014}. Moreover, single crystalline silicon (Si) exhibits a very narrowband Raman emission peak ($\rm 105~GHz$) in comparison to silica ($\rm 10~THz$), which dramatically reduces spurious broadband noise for telecom C-band applications~\cite{Lin:07}. Those features make the SOI platform appealing for further developing integrated quantum photonics devices. Notably, significant efforts were devoted to develop integrated entangled photon-pair sources (EPPS) exploiting WDM strategies~\cite{reimer_integrated_2014, jiang_silicon-chip_2015, Wakabayashi:15, grassani_micrometer-scale_2015, Rogers_2016, reimer_generation_2016, xiong_compact_2015}. So far, Si photonics based EPPS have either shown near perfect visibility ($>$~95\%) but in a single channel pair~\cite{Rogers_2016}, or, conversely, focussed on the scalibility of the source with lower visibility~\cite{reimer_generation_2016}.

This work tackles these two challenges at the same time thanks an entangled photon-pair source based on a SOI structure consisting of a micro-ring resonator coupled to a feeding waveguide. First, we optimize the ring radius in order to generate and distribute entangled photon pairs on a frequency-comb grid in the telecom C-band, symmetrically to the pump wavelength. This strategy permits exploiting the capabilities offered by DWDM devices, thus allowing to significantly increase the bit rate by using off-the-shelves standard telecom components. We optimize the frequency-comb structure linewidth as a trade-off between brightness and robustness: narrower linewidths lead to brighter sources at the cost of advanced stabilization systems~\cite{reimer_integrated_2014, rogers_twin_2015}. We successfully distribute energy-time entanglement simultaneously over multiple channel pairs using off-the-shelves DWDM components. We obtain nearly noise-free two-photon interferences pattern showing visibility figures of merit higher than 98\% in the raw data. This source can therefore serve as a pertinent standalone technological resource for large-capacity entanglement-based QKD systems.

\section{Design and optical characterizations of the photon-pair generator}
\label{sec:design}

\begin{figure}
\centering
\includegraphics[scale=0.58]{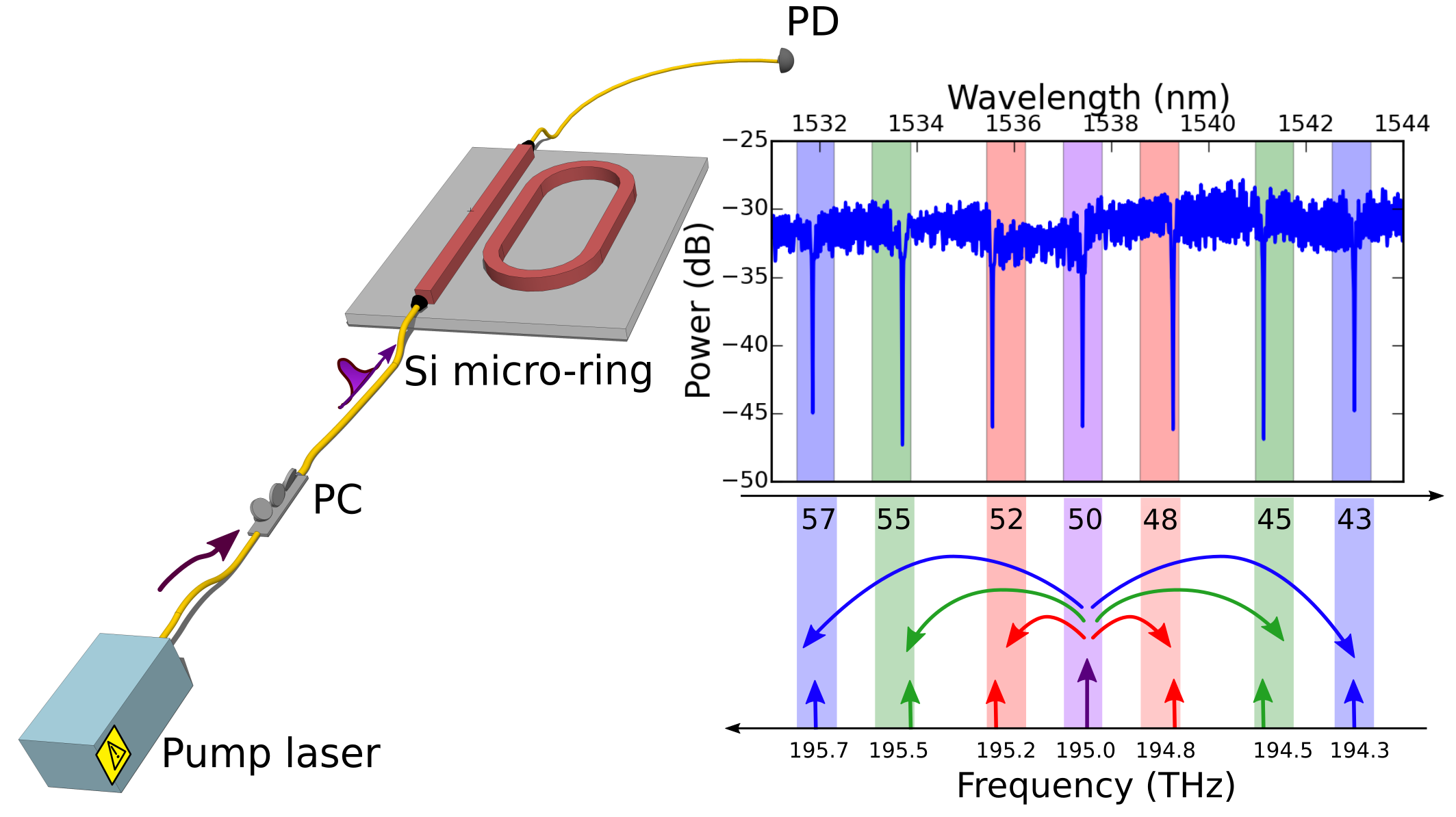}
\caption{\label{fig_1} Transmission spectrum from the sample. The left part is devoted for the setup, PC: polarization controller, PD: power detector. The right bottom part shows the strategy of our source: spectrally-correlated photon pairs are generated in ITU-grid paired channels located symmetrically with respect to the pump channel. More specifically, the pump laser is matched to a resonance peak corresponding to the international telecom union (ITU) channel 50, and twin photons are produced in two symmetrical resonance peaks, \textit{i.e}, ITU channels 48/52, 45/55, 43/57. For clarity, the most separated channel pair (ITU channels 41/59) is not illustrated. } 
\end{figure}

The silicon device employed in our experiment is a micro-ring resonator evanescently coupled to a feeding waveguide located on one side of the ring (see the left part of \figurename~\ref{fig_1}). Both ring and straight waveguides, which are etched on a SOI substrate, feature the same transverse dimensions: 600\,nm (width) by 220\,nm (height). The ring resonator radius of 60\,$\mu$m is designed to provide a moderate free spectral range (FSR) of $\sim$230\,GHz, that matches telecom channels. A coupling distance of 100\,nm has been chosen in order to maximize the energy transfer between the feeding waveguide and the micro-resonator (critical-coupling distance).

\figurename~\ref{fig_1} presents the experimental setup employed for classical characterizations. Light from a narrowband tunable telecom CW laser (Yenista Tunics) is coupled into the device. The polarization of the pump light is aligned to the TE mode of the silicon waveguide using a polarization controller (PC). Coupling the laser into and out of the chip is performed thanks to grating couplers in which light is injected by means of an almost vertical single-mode fiber. The angle is fine tuned to optimize the transmission at the pump wavelength. The total coupling losses from standard fiber to the feeding waveguide is of about 10~dB. As shown in the right side of \figurename~\ref{fig_1}, the typical transmission profile of our micro-ring resonator features a frequency-comb structure matching that of the ITU-grid. From extinction ratio measurements, we infer a quality factor (Q) of 40\,000~\cite{rogers_twin_2015}. We therefore expect the generation of photon pairs (usually referred to as signal and idler photons) symmetrically around the pump laser wavelength. The measured single-photon spectral bandwidth is 5~GHz,  corresponding to a coherence time of $\sim$100\,ps.

In order to maximize the transfer of pump light into the micro-ring resonator, we accurately study the dynamics of the frequency comb. Depending on the pump power inside the ring, the resonances are frequency-shifted by $\sim$10\,GHz/K due to thermo-optic effects in the silicon substrate~\cite{almeida_optical_2004}.
In the meantime, the power transferred to the resonator depends on the wavelength detuning between the pump laser and the shifted resonance. In general, in order to precisely determine the resonances of the ring, this thermal effect has to be addressed. Interestingly, the moderate Q factor of our resonator helps improving the robustness against such thermal drifts. Hence, a standard stabilization system using a simple temperature controller is sufficient to efficiently control the frequency-comb structure dynamics, enabling long-term stability.

\begin{figure}
\centering
 \includegraphics[scale=0.35]{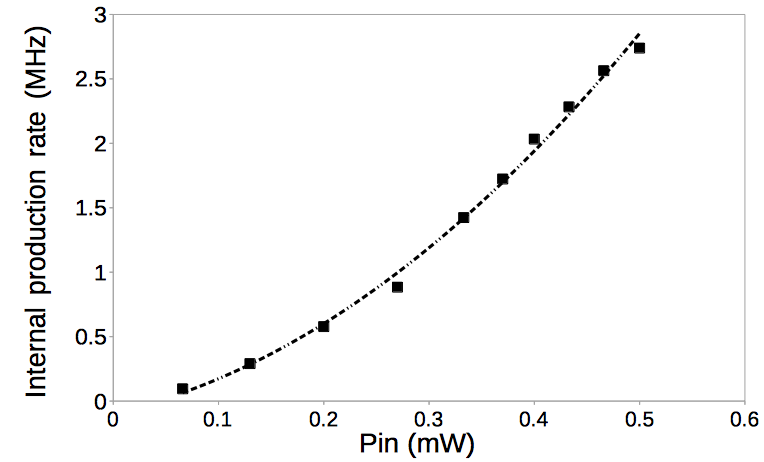}
\caption{Quadratic evolution of the internal photon-pair production rate as a function of the coupled pump power. }\label{fig_2}
\end{figure}

We now focus on the study of the non-linear process responsible for the generation of photon pairs. To this end, we fix the pump wavelength at 1537.4\,nm (ITU channel 50). Through SFWM, two pump photons are annihilated, and spectrally correlated signal and idler photons are created according to the conservation of both the energy and the momentum. The efficiency of this process is substantially enhanced by matching the pump wavelength to a resonance of the micro-ring resonator, within the SFWM gain bandwidth. Residual amplified spontaneous emission from the laser is suppressed using a tunable bandpass fiber optics filter in front of the chip (BPF, Yenista XTM-50) showing a 3-dB bandwidth of 100\,pm. Together with a series of additional narrow bandwidth BPFs, we achieve an off-band isolation of 100\,dB (see \figurename~\ref{fig_3}(a)). Another BPF of the same type is placed after the chip to isolate the emitted photons in one of the resonances. The single-photon rate is then recorded using an InGaAs avalanche photodiode (APD, IDQ-230) as a function of the injected pump power inside the feeding waveguide.
Figure.~\ref{fig_2} reports the corresponding on-chip photon-pair generation rate, taking into account all optical losses (coupling and propagation) as well as non-unity detection efficiency. The quadratic increase of the emitted rate indicates that the SFWM process is not saturated and not polluted by non-linear losses such as free-carrier absorption and two-photon absorption~\cite{engin_photon_2013}. The Q factor enables to obtain a high internal photon-pair production rate of $2\cdot 10^6$ pairs/s for $\rm 500\,\mu W$ coupled pump power only. Taking into account the narrow bandwidth of the cavity modes ($\sim$5\,GHz), we obtain a high spectral brightness of $\sim$400 pairs/s/MHz, which stands among the best values reported to date~\cite{harris_integrated_2014, azzini_classical_2012, grassani_micrometer-scale_2015, suo_generation_2015, guo_impact_2014}. We emphasize that the inherent properties of this source (quality factor, photon-pair production efficiency, resonances extinction ratio, resonance positions matching that of the ITU grid) meet the requirements for efficiently distributing entangled photon-pairs in standard telecom channels.

\section{Energy-time entanglement analysis}

\begin{figure*}[t!]
\centering
\includegraphics[scale=0.59]{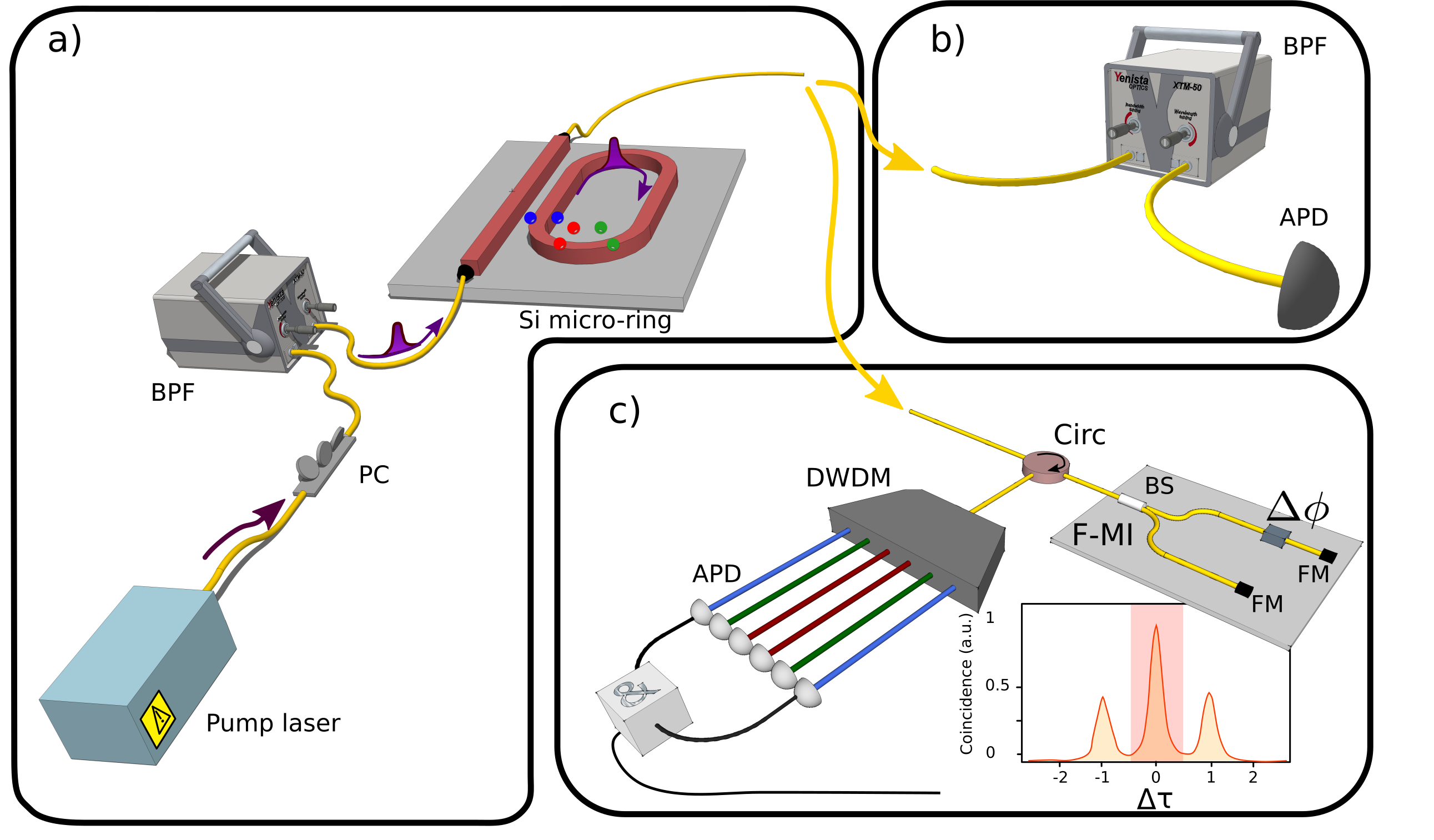}
\caption{\label{fig_3} Experimental setup. (a) Scheme of the EPPS (see main text for details). BPF: bandpass filter. (b) Employed filter acting as a spectrometer to measure the transmission profile, in the form of a frequency comb, at the output of the integrated Si-resonator. PD: power detector. (c) Coincidence histogram obtained from spectrally demultiplexed photon pairs. For the sake of clarity, the drawing has been simplified as we use two cascaded bandpass filters to reject the pump and a pair of 8x100\,GHz DWDM devices associated with a circulator and a fiber Bragg grating filter. Circ: circulator, FM: Faraday mirror, APD: avalanche photodiode, F-MI: fiber Michelson interferometer.} 
\end{figure*}

Our source has mainly been designed for entanglement-based QKD systems. As coding strategy, we consider energy-time observables which are well suited for fiber based distribution over long distances~\cite{Gisin02}. \figurename~\ref{fig_3}(a) and (c) show the entire setup for producing and analyzing energy-time entanglement.

After the paired photons are generated in the ring resonator, entanglement is revealed using a folded Franson arrangement consisting of a single unbalanced fiber Michelson interferometer (F-MI)~\cite{franson_bell_1989,Thew_Nonmax_2002}. Signal and idler photons are subsequently wavelength-demultiplexed thanks to a pair of off-the-shelves DWDM components (AC Photonics, 100\,GHz / 8-channels) which cover the ITU channels 41$\mapsto$48 (idler modes) and channels 52$\mapsto$59 (signal modes). Additional combination of broadband fiber-Bragg grating filters and DWDM permit to achieve pump laser rejection exceeding 100\,dB in all exploited channels. The generated photons are detected thanks to low-noise InGaAs APDs. The APD for the signal photons (IDQ-230) features 25\% efficiency and 250 dark counts/s, while that for the idler photons (IDQ-220) shows 20\% and 1100 dark counts/s. For both APDs, the dead-time is set to $16\,\mu s$. Both detectors are operated in the free-running regime. The transmission losses for both signal and idler photons, including output-coupling to the fiber and detection efficiencies, are measured to be 22\,dB.

Exploiting energy-time observables relies in the systematic lack of information of the pairs' creation time within the coherence time of the employed CW pump laser. In practice, twin photons pass through the unbalanced interferometer following either the same path (short-short or long-long) or different paths (long-short, or conversely)~\cite{Kaiser16}. These contributions are distinguished by measuring the arrival times of the idler photons with respect to those of the signal photons using a time interval analyzer (not represented). This enables recording a coincidence histogram comprising three peaks (see Fig.~\ref{fig_3}(c)).

To guarantee high-purity entanglement, the interferometric analysis system needs to fulfil two main requirements.
On one hand, the propagation time difference between the arms of the interferometer, $\Delta T$, has to be (i) greater than the coherence time of the single photons ($\tau\sim$ 100\,ps, deduced from the linewidth of the frequency resonances) to avoid first-order interference, and (ii) shorter than the coherence time of the CW pump laser ($\tau\sim$100\,ns) in order to have a coherent superposition between short-short and long-long contributions recorded in the central peak of the coincidence histogram~\cite{franson_bell_1989}. By isolating electronically this peak using a time window, entanglement can be revealed in the form of cross-correlation fourth-order interference.
To fulfil the two above mentioned conditions, and to take into account the detectors' timing jitter ($\sigma _{exp} = 250 \pm$ 50\,ps), the travel-time difference in the interferometer is chosen to be $\sim$350\,ps.

On the other hand, long-term interferometer phase stability is ensured using an active system based on a dither loop. To this end, a 1560.5\,nm reference laser (RIO Orion) is sent to the interferometer through a dense wavelength division multiplexer in the counter-propagating direction compared to that of the entangled photons. The relative phase between the two arms of the interferometer is monitored via an intensity measurement and a piezoelectric fiber stretcher placed in its the long arm corrects unwanted drifts. Our stabilization scheme has a loop bandwidth of 300\,Hz which is fast enough to guarantee a stability of $\Delta \phi < \frac{2\pi}{50}$. The two-photon phase is tuned by changing the wavelength of the reference laser through temperature control.

\begin{figure*}[t!]
\begin{center}
\begin{tabular}{c c c c}
(a) & & (b) & {}\\
(c) & \includegraphics[scale=0.18]{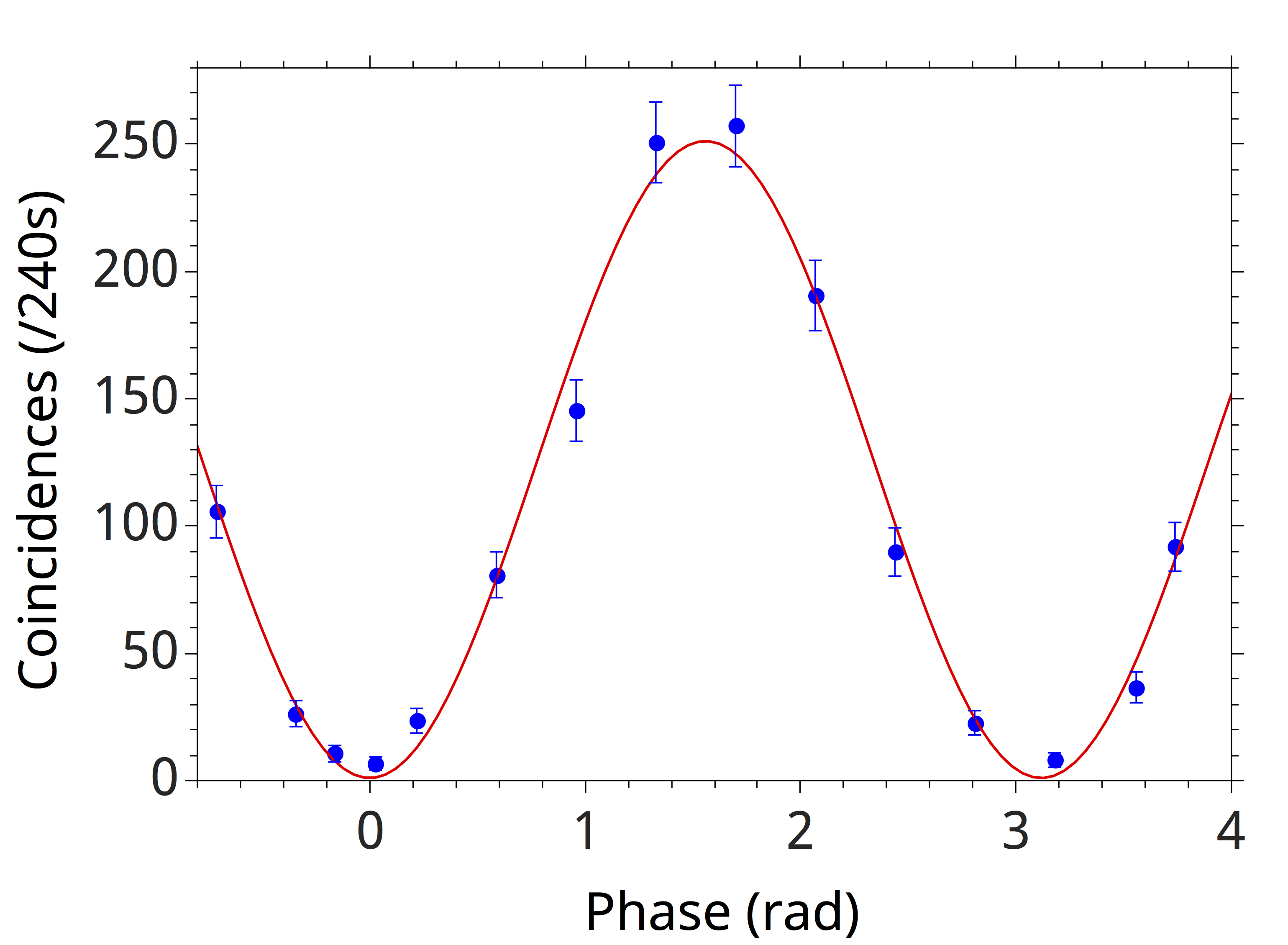} & (d) & \includegraphics[scale=0.18]{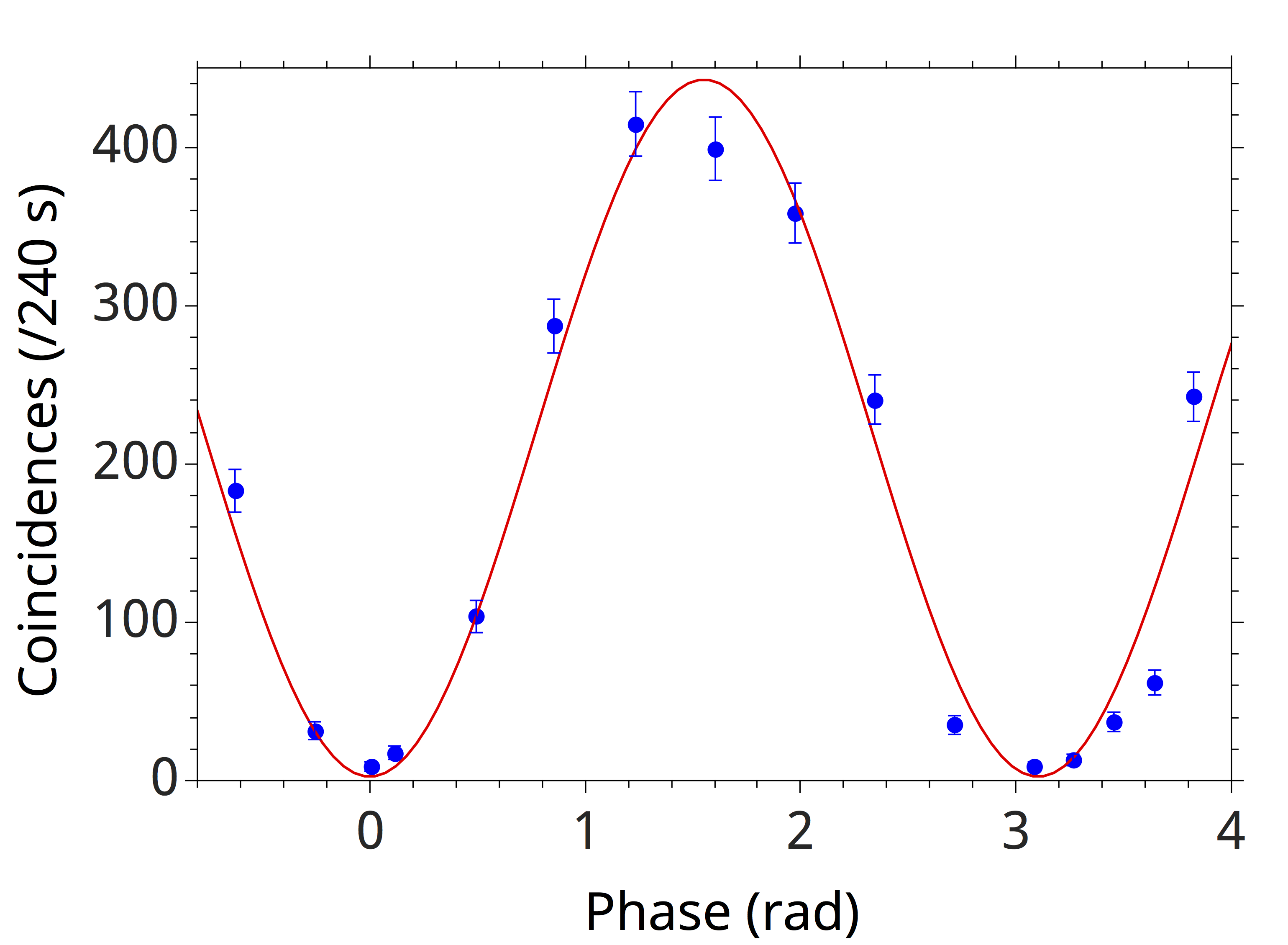} \\
{} & \includegraphics[scale=0.18]{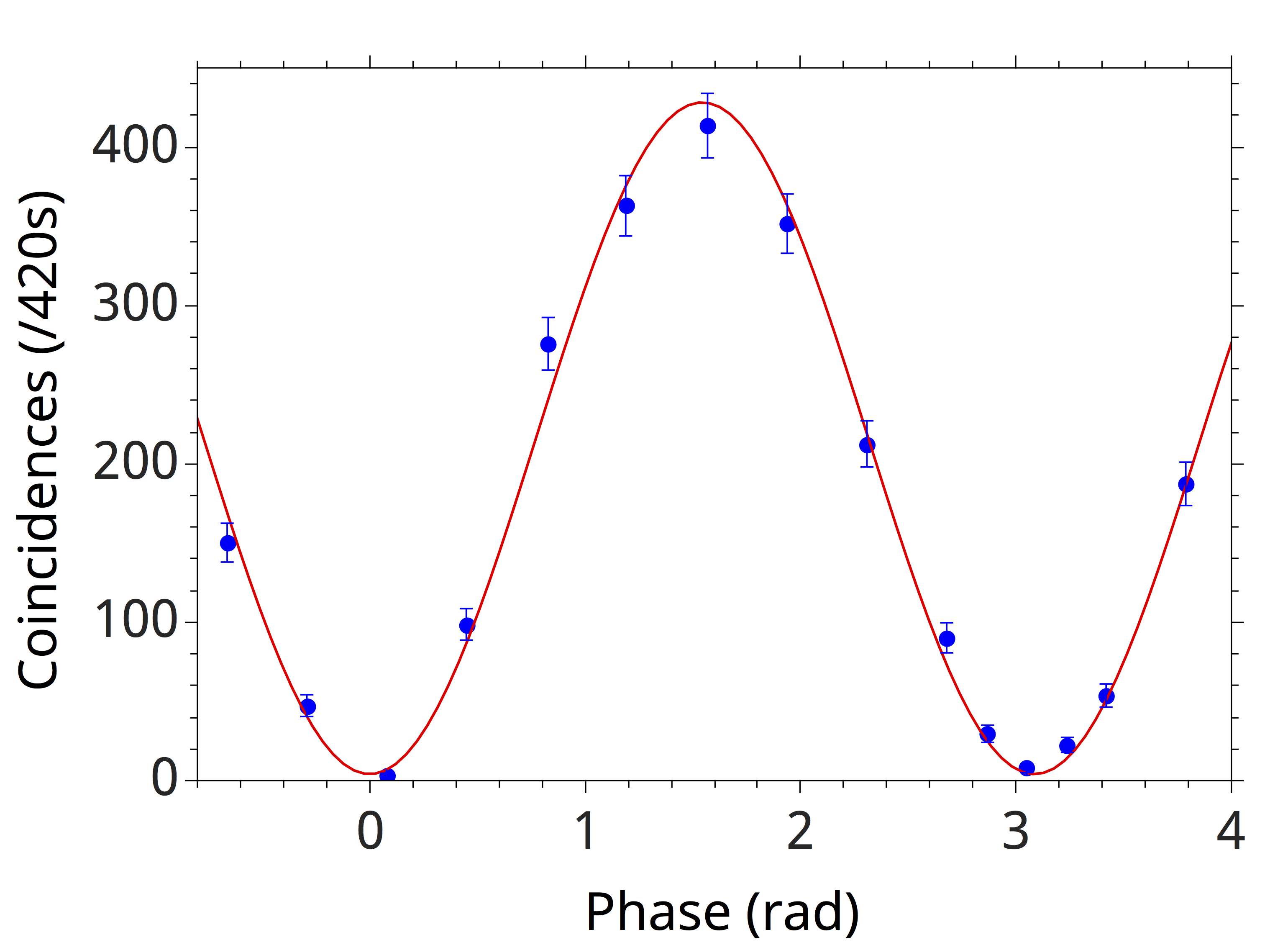} & & \includegraphics[scale=0.18]{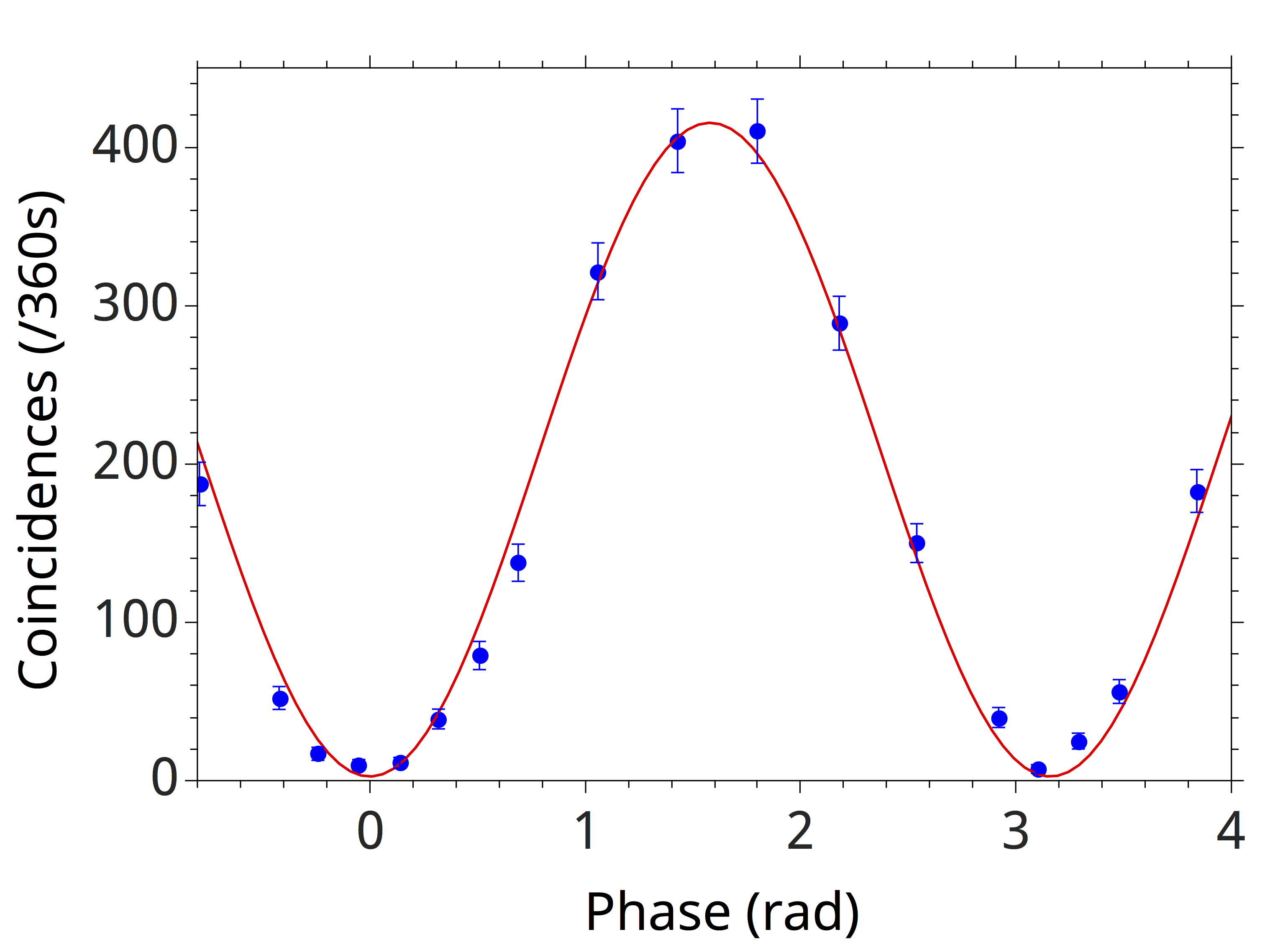}\\
\end{tabular}
\begin{tabular}{c c c c}
(e) & & & {}\\
& \includegraphics[scale=0.32]{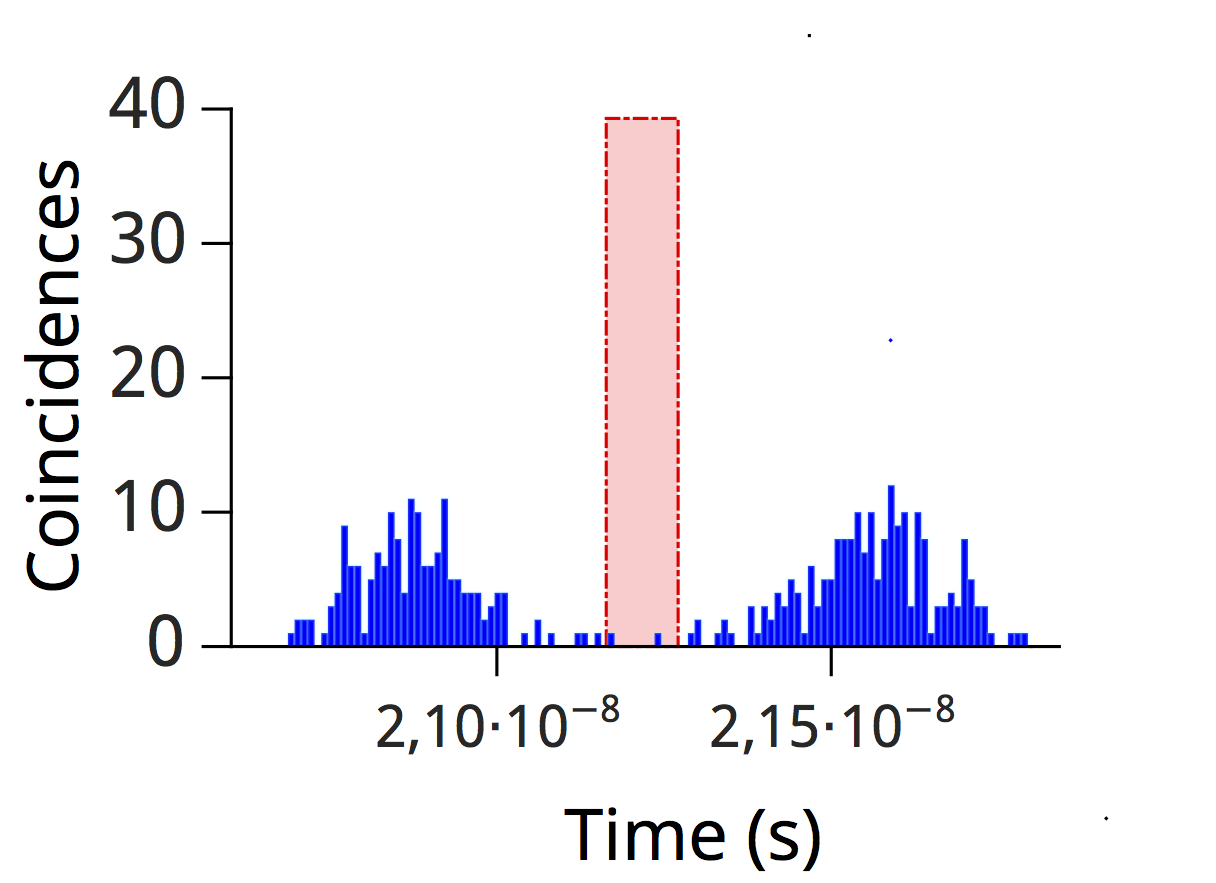} & \includegraphics[scale=0.32]{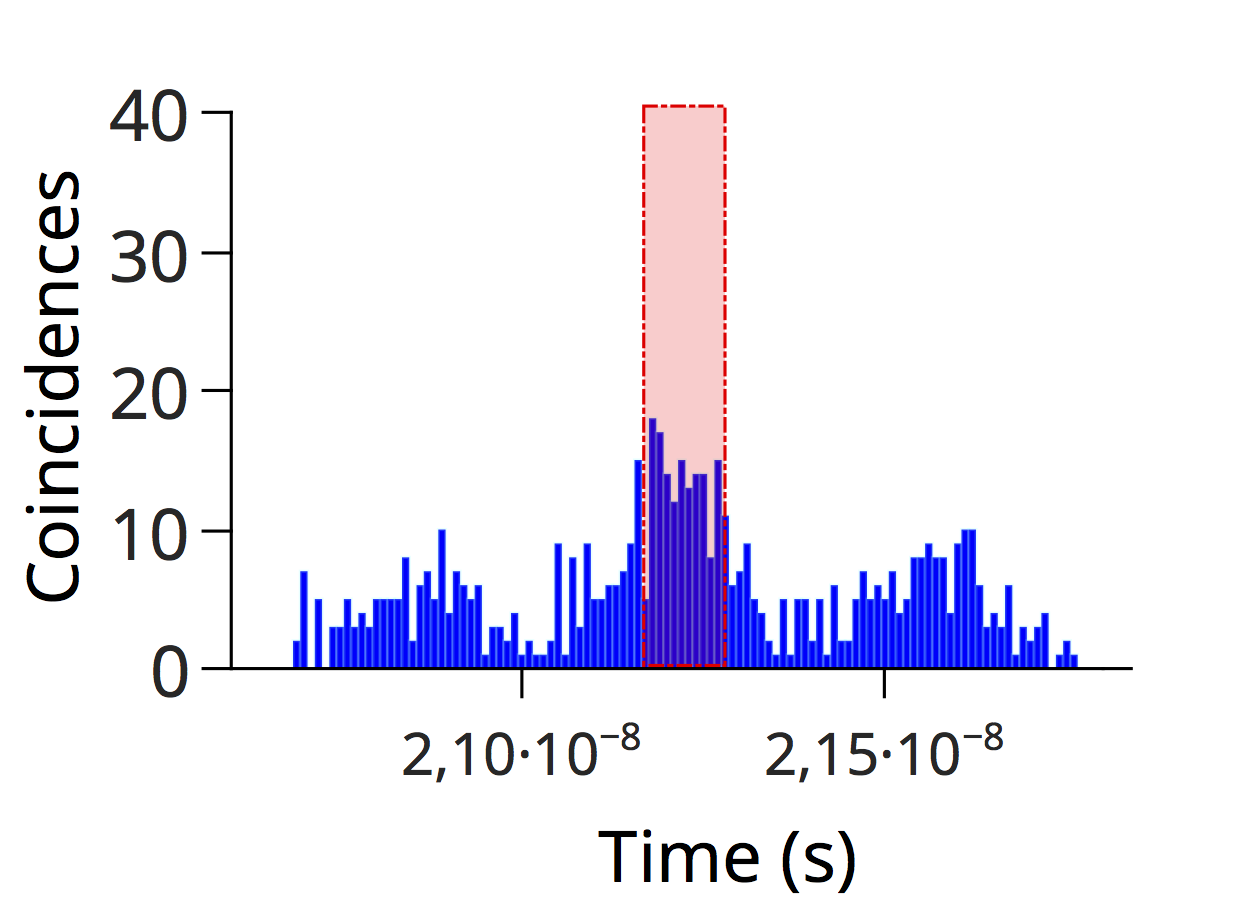} & \includegraphics[scale=0.32]{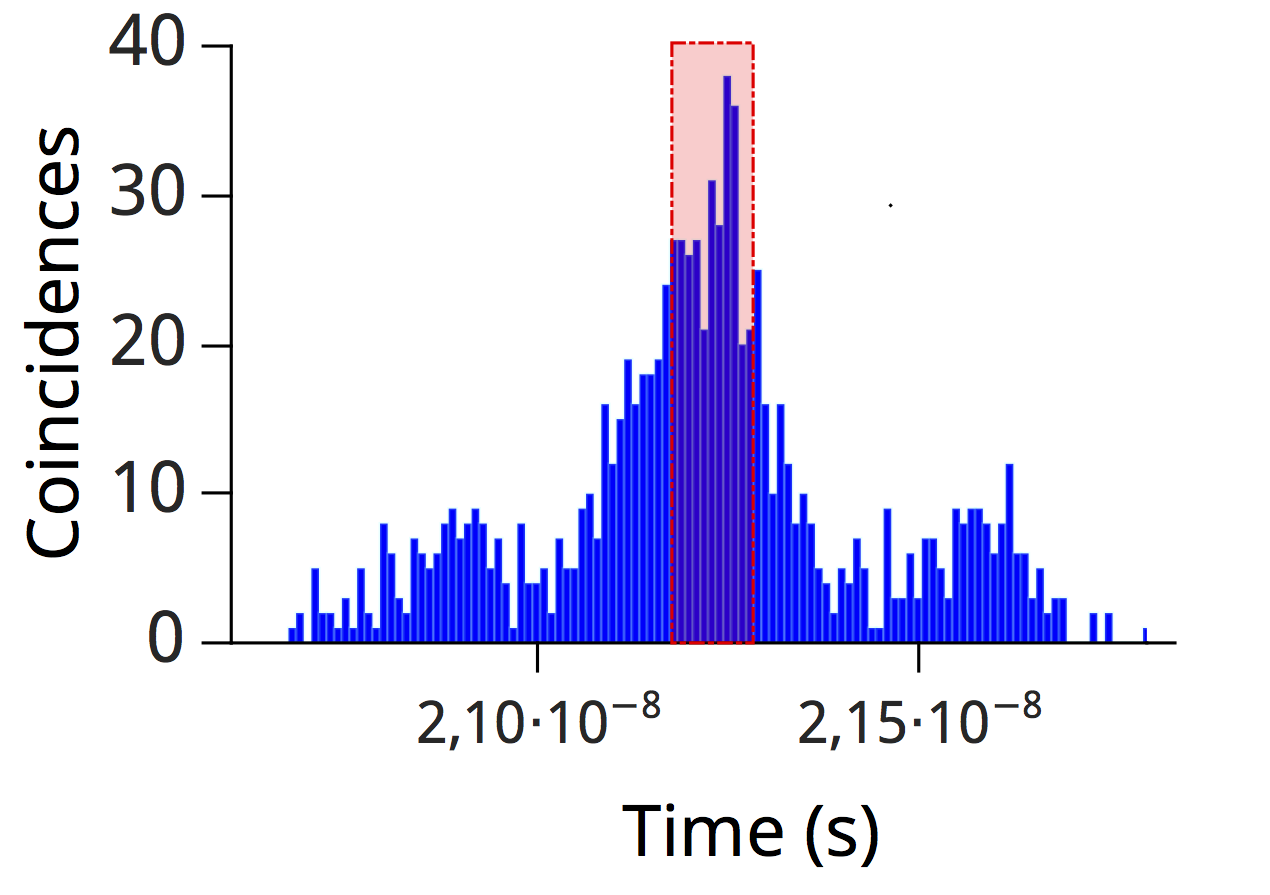}\\
 \end{tabular}
\caption{\label{fig_4} Two-photon interference fringes obtained with energy-time entangled photons. The curves show the coincidence counts for the ITU paired channels 48/52 (a) 45/55 (b) 43/57 (c) and 41/59 (d) as a function of the interferometer phase. Experimental data are assumed to have an error-bar magnitude of $\sqrt {N}$ (\textit{N} stands as the number of coincidences measured) as the photon pair emission statistic is poissonian. (e) Raw histograms corresponding to 3 particular phase values for curve (a): minimum, half a maximum, and maximum of the two-photon interference fringe.}
\end{center}
\end{figure*} 

We now analyse the coincidence events registered in the central peak, which amounts to a sinusoidal oscillation as a function of the two-photon phase, $\Delta \phi$, \textit{i.e.} the sum of the phases acquired by the two individual photons in the interferometer.
We exploit the channel capacity offered by our DWDM devices to measure two-photon interference fringes between various paired channels. To this end, we subsequently realize this measurement in the ITU paired channels 48/52, 45/55, 43/57, and 41/59, all symmetrically located on both sides of the pump channel (ITU 50). We obtain sinusoidal modulations of the coincidence rates as shown in \figurename~\ref{fig_4}(a)--(d). Moreover, \figurename~\ref{fig_4}(e) shows raw coincidence histograms corresponding to three particular phase values: a minimum, half a maximum and a maximum. The quality of the measured entangled state is assessed by measuring the two-photon interference fringe visibilities, defined as $C=(C_{max}-C_{min})/(C_{max}+C_{min})$. Here, $C_{max}$ and $C_{min}$ denote the maximum and the minimum coincidence rates, respectively, obtained from fitting the data with a sinusoidal function. We infer visibilities of 99.2\,$\pm$\,2.3\%, 98.9\,$\pm$\,2.7\%, 98.1\,$\pm$\,0.9\%, 98.8\,$\pm$\,1.5\%, respectively, using only free fit parameters.
When subtracting the accidental coincidences originating only from the detectors' dark counts, we obtain net visibilities of 99.7\%, 99.4\%, 98.6\%, 99.3\%, respectively. Note that due to an non optimized path length difference  in the interferometer, the 3 peaks of the coincidence histogram are not perfectly separated. We compute a 0.3 \% degradation in the visibilities because of this crosstalk. Those extremely high quantum interference visibilities stand as a clear witness of the non-classical correlations existing between the paired photons, as they not only exceed exceed largely the threshold of $1/\sqrt{2}=70.7$\% but also are very close to unity~\cite{grassani_micrometer-scale_2015, franson_bell_1989,Kaiser16}. Let us stress that these results stand as the highest raw quantum interference visibilities for energy-time entangled photon pairs for micro/nanoscale EPPSs \cite{Wakabayashi:15, PhysRevApplied.4.021001, silverstone_qubit_2015, silverstonej._w._-chip_2014, guo_impact_2014, Rogers_2016, Kumar_2015}. We also emphasize that this nearly perfect visibilities demonstrate both the high quality of the Si micro-ring EPPS and the ability of our chip to produce high rates of entangled photon pairs without background excess noise. Note that obtaining such a level of visibility has been made possible by paying particular attention for reducing well-known noise contributions: (i) dark counts are limited thanks to the use of very low-noise APDs, (ii) low pump powers (few 100\,$\mu W$) are exploited for ensuring negligible multiple photon-pair events, and (iii) a short input fiber pigtail is used for minimizing Raman-scattering.

\section{Conclusion and outlook}

We have reported the conception, realization, and full characterization of a telecom compliant EPPS based on a silicon micro-ring resonator chip showing as well compactness and a high brightness ($\sim$400\,pairs/s/MHz for $\rm 500\,\mu W$ coupled pump power). The near-perfect two-photon interference-pattern visibilities obtained in different paired channels clearly demonstrate the ability of our sources to stand as a key resource for future entanglement-based QKD systems, implemented over a large number of channels in the telecom C-band. We strongly believe that our proof of principle device sets the conceptual basis for further developments, such as exploiting DWDM components with higher spectral capacity. We anticipate the extension of our EPPS over the entire telecom C-band, relying on broadband phase matching condition engineering, as it has recently been demonstrated in a different technology platform~\cite{reimer_generation_2016}. In this perspective, one has to tailor the waveguide dimensions to obtain small and anomalous chromatic dispersion over the bandwidth of interest~\cite{ferrera_low_2009}. Finally, the generation of multiple photon-pairs makes readily available multipartite entanglement and then extends significantly the ability of integrated quantum photonics~\cite{pysher_parallel_2011}.

\section*{Funding}
Agence Nationale de la Recherche (\#{}ANR-SITQOM-15-CE24-0005); European Commission FP7-ITN (\#{}PICQUE-608062); 
Erasmus+ program (\#{}2014-1-IT02-KA103-000041).

\section*{Acknowledgment}
The authors thank O. Alibart, D. Bonneau, and V. D'Auria for fruitful discussions, as well as P. Bassi for his support and inputs. The authors also acknowledge technical support from IDQ and Yenista.

\end{document}